\documentclass[%
pre,
 aps,%
 amsmath,amssymb,
reprint,%
]{revtex4-1}

\usepackage{color}
\usepackage{bm}
\usepackage{amsmath}
\usepackage{amssymb}
\usepackage{graphicx}

\begin{document}

\newtheorem{lemma}{Lemma}
\newtheorem{corollary}{Corollary}

\title{Internal wave pressure, velocity, and energy flux from density perturbations}

\author{Michael R. Allshouse}
\email{mallshouse@chaos.utexas.edu}
\affiliation{Center for Nonlinear Dynamics and Department of Physics,${\  }$University of Texas at Austin, Austin, TX 78712, USA}
\author{Frank M. Lee}
\affiliation{Institute for Fusion Studies and Department of Physics, University of Texas at Austin, Austin, TX 78712, USA}
\author{Philip J. Morrison}
\affiliation{Institute for Fusion Studies and Department of Physics, University of Texas at Austin, Austin, TX 78712, USA}
\author{Harry  L.  Swinney}
\affiliation{Center for Nonlinear Dynamics and Department of Physics, University of Texas at Austin, Austin, TX 78712, USA}

\begin{abstract}

Determination of energy transport is crucial for understanding the energy budget and fluid circulation in density varying fluids such as the ocean and the atmosphere.  However, it is rarely possible to determine the energy flux field $\bm{J} = p \bm{u}$, which requires simultaneous measurements of the pressure and velocity perturbation fields, $p$ and $\bm{u}$.  We present a method for obtaining the instantaneous $\bm{J}(x,z,t)$ from density perturbations alone: a Green's function-based calculation yields $p$, and $\bm{u}$ is obtained by integrating the continuity equation and the incompressibility condition.  We validate our method with results from Navier-Stokes simulations:  the Green's function method is applied to the density perturbation field from the simulations, and the result for $\bm{J}$ is found to agree  typically to within 1\% with $\bm{J}$ computed directly using $p$ and $ \bm{u}$ from the Navier-Stokes simulation.
We also apply the Green's function method to density perturbation data from laboratory schlieren measurements of internal waves in a stratified fluid, and the result for $\bm{J}$ agrees to within $6\%$ with results from Navier-Stokes simulations.  Our method for determining the instantaneous velocity, pressure, and energy flux fields applies to any system described by a linear approximation of the density perturbation field, e.g., to small amplitude lee waves and propagating vertical modes.  The method can be applied using our Matlab graphical user interface {\it EnergyFlux}.\\

\end{abstract}

\maketitle



\section{Introduction}\label{sec:Introduction}


The transport of energy in the ocean by internal gravity waves is vital for thermohaline circulation, ocean mixing, and the ocean's overall energy budget~\cite*{munk98,alford03,wunsch04}. The rate at which internal wave energy is transported through an area is given by the baroclinic energy flux, 
\begin{equation}
\bm{J} = p \bm{u},
\end{equation}
where $p$ is the pressure perturbation from the hydrostatic background and $\bm{u}$ is the velocity perturbation from the background flow.  For periodic internal waves, the energy flux is often averaged over a period of the internal wave, though this precludes application to aperiodic disturbances such as internal solitary waves.  In theoretical analyses~\cite*{balmforth02, smith03, balmforth09} and numerical simulations~\citep{lamb04, niwa04, zilberman09, king09, king10, gayen10, gayen11, rapaka13}  the pressure perturbation and velocity fields are known, making the calculation of the energy flux straightforward. 

However, in laboratory studies the pressure perturbation field is difficult to measure directly, and obtaining the velocity field requires a second simultaneous measurement.  In tank-based experiments the time-averaged energy flux has been calculated using data from the velocity or density fields, as reviewed in section \ref{sec:background}.  Here, using a Green's function approach, we present a more generally applicable method for calculating the instantaneous pressure, velocity, and energy flux from the density perturbation field; thus the method can be applied to both periodic and aperiodic data.  The method was developed for use on laboratory density perturbation data but should also be applicable to field observations.   

This paper is organized as follows. Section~\ref{sec:background} reviews approaches developed for calculating the energy flux for time-periodic data. Section~\ref{sec:Theory} presents the derivation of our method for calculating the instantaneous energy flux field $\bm{J}$.  In subsection~\ref{sec:genTheory} we start with the linear Euler's equations and derive expressions for the pressure perturbation and the two velocity components in terms of the density perturbation. These allow for a general expression for $\bm{J}$ in terms of the density perturbation field. In subsection~\ref{sec:analytical} a Green's function method is used to solve for the pressure perturbation field from a density perturbation field, which will be given by synthetic schlieren data. Section~\ref{sec:Methods} describes our numerical simulations and experiments and compares their results. In subsection~\ref{subsec:full}, our method is verified by comparing results for $\bm{J}$ calculated from a simulated density perturbation field with results obtained directly from numerical simulations. Subsection~\ref{subsec:Labdata} presents the results of applying the method to laboratory data taken in a portion of the domain. Finally,  section~\ref{sec:Conclusions} presents our conclusions and discusses broader applications of our method.  To aid in applying this method, we have developed a Matlab GUI, {\it EnergyFlux}, which is discussed in the appendix and provided in the supplementary materials.
  

\section{\label{sec:background} Background}


Previously, the energy flux has been computed from velocity data by two different approaches (subsection~\ref{subsec:vel-based-flux}), and from density perturbation data by two additional approaches (subsection~\ref{subsec:density-based-flux}). These four approaches provide leading order approximations for the time-averaged energy flux from measurements, but differ from our approach in that they cannot capture transient features because they rely on periodicity in time.

\subsection{Velocity-based energy flux approaches} ~\label{subsec:vel-based-flux}

The velocity-based approaches for calculating the energy flux use continuity, incompressibility, and the linear Euler's equations, with the assumption of time-periodic internal waves. These approaches obtain the energy flux in terms of the stream function~\citep{balmforth02,lee14}, obviating the need for the pressure field. The two velocity-based approaches differ in how they calculate the stream function from velocity data: the first approach uses modal decomposition, while the second obtains the stream function using path integrals. 

The approach that makes a modal decomposition of the velocity field assumes hydrostatic balance (requiring the forcing frequency to be much smaller than the buoyancy frequency)~\citep{petrelis06}.
An application of this approach to a tank-based experiment by~\citet{echeverri09} dropped the hydrostatic balance requirement and added a viscous correction. Most of the energy they observed was contained in the first mode, and the energy flux in modes higher than three was not measurable.

The modal-decomposition approach assumes time periodicity in obtaining the time-averaged energy flux. A periodic signal is obtained using Fourier transforms, but the accuracy is limited because typical data records are only a few periods long, and also nonlinearities can lead to energy transfer to other frequencies. Further, a modal analysis requires determining the shapes of the vertical modes, but density data spanning the entire fluid depth are often not available.  Also, in laboratory experiments the high viscous dissipation limits the results to only the first few modes. 

The second velocity-based approach avoids modal decomposition and calculates the stream function directly~\citep{lee14}.  Instantaneous velocity fields obtained by particle image velocimetry are used to obtain the stream function.  By calculating multiple path integrals between a base point and each point in the domain, this approach averages out some of the noise inherent to experimental measurements; however, accurate results depend on the base point of the integration being either at the boundary of the system,  where the stream function is zero, or in a region of the domain where the velocity vanishes.  While this approach also relies on time-periodicity of the field, a more complete representation of the stream function is possible compared to the first approach.

\subsection{Density-perturbation-based energy flux approaches}~\label{subsec:density-based-flux}

The first approach that uses the density perturbation field is that of \citet{nash05}, who obtained the energy flux from observational oceanic data for density in a water column.  The density perturbation is assumed to be the only contribution to the pressure perturbation, and thus integration of the density perturbations results in the hydrostatic pressure perturbations.  This assumption is valid when the buoyancy frequency of the ocean is much larger than the tidal frequency.  The velocity perturbation used in this approach removes the mean time-periodic background velocity and a constant to satisfy the baroclinicity assumption.  In regions of the ocean where the most active internal wave fields exist, the time-averaged energy flux has been measured with this approach and used to verify corresponding ocean modeling~\citep{rudnick03, alford15}.

The approach of \citet{nash05} can be applied not only to ocean measurements but also to laboratory measurements if synthetic schlieren and particle image velocimetry are performed simultaneously, as was done by \citet{jia14}. However, the approach requires both density and velocity data for the entire water column. Additionally, the calculation of the pressure perturbations assumes that there is no contribution from the dynamic pressure, which is reasonable for oceanic data given the slow time scale over which the velocity field changes, but this assumption is invalid for some laboratory experiments and also in ocean settings where the water column is weakly stratified.  

A second approach that uses the density perturbation field relies on Boussinesq polarization relations and eigenvector solutions of the linear and inviscid internal wave equations.  The polarization relations, which assume periodic flows and plane wave solutions, provide a direct link between the amplitude and phase of any of the velocity components, density perturbation, pressure perturbation, and vertical isopycnal displacement~\citep{sutherland10}.  These relationships are functions of the internal wave frequency.  The strength of this approach is that given a periodic or nearly periodic flow, a determination of the velocity field through PIV or isopycnal displacement (using synthetic schlieren) can be used to obtain the pressure and density fields~\citep{clark10}.  When the flow field is not strictly periodic but is dominated by a single frequency, spectral methods can be used to decompose the system into its modal contributions, and the polarization relations can be applied to each modal component.  This approach provides a direct means for calculating the pressure and thus the energy flux, but the approach relies on accurate spectral decomposition of the fields.

The polarization approach was applied to synthetic schlieren measurements of the isopycnal displacement field by \citet{clark10}, who investigated internal wave beams radiating away from a turbulent patch.  To determine the dominant wave frequency and wavenumber, multiple transects normal to the generated beams over multiple periods were analyzed using FFT methods.  Then the maximum displacement amplitude based on the spatially averaged envelope was calculated.  Combining these two results with the polarization relations yielded the energy flux generated at the dominant frequency and wavenumber pair. 

While the approach of \citet{clark10} provides a notable first step for obtaining energy flux from synthetic schlieren data, it has some limitations.  First, it requires that the system be periodic or nearly periodic.  In an aperiodic or transient flow field, the polarization relations require a large number of frequency-wavenumber pairs to  reproduce the flow field.  The necessity of accurate modal decomposition in both space and time of the synthetic schlieren data makes the averaging process difficult~\citep{clark10}.  Another limitation is that the spatial averaging along the beam assumes no viscous dissipation, while the dissipation can be significant for laboratory internal waves~\citep{lee14}.  


\section{Theory} \label{sec:Theory}


Our approach uses the density perturbation field to calculate the instantaneous pressure, velocity, and energy flux fields.  Starting with the linear Euler's, continuity, and incompressibility equations, we derive expressions for the pressure and velocity perturbation fields in terms of the density perturbation field.  Section~\ref{sec:genTheory} presents these relationships without assuming any particular form for the buoyancy frequency $N$. For the specific case of uniform $N$, a solution for the pressure perturbation field is found in terms of the density perturbation field in section~\ref{sec:analytical}.  

\subsection{Energy flux from a density perturbation field} \label{sec:genTheory}

To calculate the energy flux from the density perturbation field, the pressure and velocity must first be obtained in terms of the density perturbations. Assuming inviscid flow, we start with the two-dimensional Euler's equations, which give the linear wave equations that are the foundation of our approach.  We obtain a partial differential equation that gives the pressure perturbations instantaneously from the density perturbation field, which acts as a source term, and then the incompressibility and the continuity equations together yield both velocity components as functions of the density perturbations.

The linearized two-dimensional Euler's equations for the density $\rho_0(z) + \rho(x,z,t)$ and pressure $p_0(z) + p(x,z,t)$, where $\rho_0(z)$ and $p_0(z) $ are in hydrostatic balance, and  the velocity $\bm{u}(x,z,t)$ are: 

\begin{eqnarray}
&&\frac{\partial u}{\partial t} = -\frac{1}{\rho_0} \frac{\partial p}{\partial x}\,,  
\qquad
\frac{\partial w}{\partial t} = -\frac{1}{\rho_0} \frac{\partial p}{\partial z} - \frac{\rho}{\rho_0} g\,, \label{dudt}    \\
&&
\frac{\partial \rho}{\partial t} = \frac{N^2 \, \rho_0}{g} w\,,  \qquad
\frac{\partial u}{\partial x} + \frac{\partial w}{\partial z} = 0\,,
 \label{div1}
\end{eqnarray}
where $g$ denotes the gravitational acceleration, 
$x$ and $z$ are the horizontal and vertical  coordinates, respectively, $u$ and $w$ are the corresponding components of the velocity $\bm{u}$, and the buoyancy frequency $N$ is given by
\begin{align}
N^2 = -\frac{g}{\rho_0} \frac{d \rho_0}{d z}.
\label{N}
\end{align}
The energy density is given by
\begin{align}
E =& \frac{\rho_0}{2} ( u^2 + w^2 ) - \frac{\rho^2 g}{ 2 \, d \rho_0 / d z} \label{energy},
\end{align}
which together with  the energy flux $\bm{J}$ satisfies conservation of energy, 
\begin{align}
\frac{\partial E}{\partial t} + \nabla \cdot \bm{J} =& 0 \label{encons}.
\end{align}
Using the equations of motion \eqref{dudt} and \eqref{div1},  we have the energy flux from \eqref{encons},
\begin{align}
\bm{J} = u p \bm{\hat{x}} + w p \bm{\hat{z}}\,, \label{J}
\end{align}
which is the main object of our consideration.

Next, using  \eqref{dudt} to obtain the time derivative of $\nabla\cdot\bm{u}$ yields  
\begin{equation}
\begin{split}
\frac{\partial}{\partial x} \frac{\partial u}{\partial t} + \frac{\partial}{\partial z} \frac{\partial w}{\partial t} = \\
&\phantom{=} \hspace{-2cm} \frac{\partial}{\partial x} \left( -\frac{1}{\rho_0} \frac{\partial p}{\partial x} \right) + \frac{\partial}{\partial z} \left( -\frac{1}{\rho_0} \frac{\partial p}{\partial z} - \frac{\rho}{\rho_0} g \right)=0\,
\end{split}
 \label{divacc}
\end{equation}
which upon rearranging gives the following partial differential equation:
\begin{align}
\frac{\partial^2 p}{\partial x^2} + \frac{\partial^2 p}{\partial z^2} + \frac{N^2}{g} \frac{\partial p}{\partial z}
=& - N^2 \rho  -  g \frac{\partial \rho}{\partial z}. 
\label{peqn}
\end{align}
Equation \eqref{peqn}, together with boundary conditions discussed in section \ref{sec:analytical},  yields the  pressure perturbation field from a source that is determined by the density perturbation field at any given instant in time. We denote the solution of \eqref{peqn} by the functional $p[\rho]$. 

To obtain a  more intuitive and easier-to-solve equation, we transform \eqref{peqn} to a standard form in terms of a new variable $q$:
\begin{align}
p(x,z) = q(x,z) \exp{ \left[ -\frac{1}{2g} \int^{z}{dz' N^2(z')}\right] }\,.
\label{pressure}
\end{align}
The relationship between $q$ and $\rho$ is then
\begin{align}
\begin{split}
\frac{\partial^2 q}{\partial x^2} + \frac{\partial^2 q}{\partial z^2} - \left( \frac{N}{g} \frac{\partial N}{\partial z} + \frac{N^4}{4g^2} \right) q  = \\
&\phantom{=} \hspace{-5cm} - \left( N^2 \rho  +  g \frac{\partial \rho}{\partial z} \right) \exp{ \left[ \frac{1}{2g} \int^{z}{dz' N^2(z')}\right] }\,,
\end{split}
 \label{qeqn}
\end{align}
which when solved gives $p[\rho]$ via \eqref{pressure}.

The vertical component of the velocity $w$ is given by rearranging \eqref{div1}, 
\begin{align}
w = \frac{g}{N^2 \rho_0} \frac{\partial \rho}{\partial t}\,.
 \label{w}
\end{align}
Using  $w$, we find the horizontal component of the velocity $u$ from the  incompressibility condition by integrating in $x$, 
\begin{align}
u = - \int^{x} \!\!\! dx\, \frac{\partial w}{\partial z}  = - \int^{x} \!\!\!  dx \,  \frac{\partial}{\partial z} \left( \frac{g}{N^2 \rho_0} \frac{\partial \rho}{\partial t} \right) \,.
 \label{u}
\end{align}
The integration constant is zero if we take the initial point of integration to be at a location where the horizontal velocity is known to be zero. 

Finally, using \eqref{w} and \eqref{u}, we obtain the desired result, the instantaneous energy flux \eqref{J} entirely in terms of the density perturbation field $\rho$, provided we know $p[\rho]$, the solution of   \eqref{peqn}  for the  pressure perturbation field,

\begin{align}
\begin{split}
\bm{J}(x,z,t) = - p[\rho] \, g \int^{x}\!\!\!  dx \, \frac{\partial}{\partial z} \left( \frac{1}{N^2 \rho_0} \frac{\partial \rho}{\partial t} \right)   \bm{\hat{x}} + \\ &\phantom{=} \hspace{-1.25cm} \frac{p[\rho] \, g}{N^2 \rho_0} \frac{\partial \rho}{\partial t}\, \bm{\hat{z}} .
\end{split}
 \label{J2}
\end{align}

\subsection{Green's function approach for uniform N} 
\label{sec:analytical}

Before solving \eqref{qeqn} for the pressure perturbations, the boundary conditions must  be specified.   A detailed discussion of the experimental setup will be given in section \ref{sec:expt}, but for now we note that our boundary conditions are for a domain that will represent laboratory data taken from a tank where the top and bottom boundaries are visible, while the left and right boundaries are not, because they are taken to be far away.  As an approximation  of our laboratory domain, periodic boundary conditions are assumed for the left $(x=0)$ and right $(x=l)$ boundaries, and no-flux boundary conditions are assumed for the top $(z=0)$ and bottom $(z=h)$ of the domain.   The periodic boundary conditions for the  horizontal direction are reasonable since disturbances do not sense the actual boundary in that direction, while the  no-flux conditions in the vertical direction are appropriate since the top and bottom boundaries of the measurement window are the solid boundary of the tank and the free surface.

The boundary conditions required for solving \eqref{peqn} follow from force balance.  For the horizontal periodic boundary conditions,  the first equation of  \eqref{dudt}  implies  
\begin{align}
\frac{\partial p}{\partial x} \bigg|_{x=0} = \frac{\partial p}{\partial x} \bigg|_{x=l} \,.
\label{pbdy1}
\end{align}
Similarly, applying the no-flux boundary condition on the top and bottom boundaries requires the vertical velocity there to be zero for all time, and this implies zero vertical force there as well. Then the second equation of  \eqref{dudt} gives
\begin{align}
\left( \frac{\partial p}{\partial z} - \frac{\rho}{\rho_0} g \right) \bigg|_{z=0} = \left( \frac{\partial p}{\partial z} - \frac{\rho}{\rho_0} g \right) \bigg|_{z=h} = 0. \label{pbdy2_0}
\end{align}
However, the first equation of \eqref{div1} tells us that the density perturbation does not change with time at the top and bottom boundaries since the vertical velocity is zero there. Since initially the density perturbation on those boundaries is zero, it remains zero for all time. Thus \eqref{pbdy2_0} gives the following boundary condition for the top and bottom boundaries:
\begin{align}
\frac{\partial p}{\partial z} \bigg|_{z=0} = \frac{\partial p}{\partial z} \bigg|_{z=h} = 0. \label{pbdy2}
\end{align}
Because of the transformation \eqref{pressure}, the boundary conditions on $p$, \eqref{pbdy1} and \eqref{pbdy2}, imply the following boundary conditions on the variable $q$:

\begin{equation}
\begin{split}
\frac{\partial q}{\partial x} \bigg|_{x=0} &= \frac{\partial q}{\partial x} \bigg|_{x=l}  \\
\frac{\partial q}{\partial z} - \frac{N^2}{2g}q \, \bigg|_{z=0} &= \frac{\partial q}{\partial z} - \frac{N^2}{2g} q \, \bigg|_{z=h} = 0. \label{qbdy}
\end{split}
\end{equation}

In this section we consider the case where the buoyancy frequency profile is taken to be uniform, $N=N_0$\textcolor{red}.  For such a profile, the equation for the pressure perturbation field \eqref{peqn} and the boundary conditions \eqref{qbdy} simplify to give
\begin{align}
\frac{\partial^2 q}{\partial x^2} + \frac{\partial^2 q}{\partial z^2} - \frac{N_0^4}{4g^2} q  = - f(x,z), \label{qeqnN0}
\end{align}
\begin{equation}
\begin{split}
\frac{\partial q}{\partial x} \bigg|_{x=0} &= \frac{\partial q}{\partial x} \bigg|_{x=l}, \\ \frac{\partial q}{\partial z} - \frac{N_0^2}{2g} q \, \bigg|_{z=0} &= \frac{\partial q}{\partial z} - \frac{N_0^2}{2g} q \, \bigg|_{z=h} = 0\,,
 \label{qbdyN0}
\end{split}
\end{equation}
where
\begin{align}
f(x,z) = \left( N_0^2 \rho  +  g \frac{\partial \rho}{\partial z} \right) \exp{ \left( \frac{N_0^2}{2g} z\right) }. 
\label{f}
\end{align}

Next, the variables $q$ and $f$ are Fourier expanded in the horizontal direction, 
\begin{equation}
\begin{split}
q(x,z) &= \mathrm{Re} \, \bigg\{ \sum\limits_{k} {Q_k(z) e^{- i k x/l}} \bigg\} \\
f(x,z) &= \mathrm{Re} \, \bigg\{ \sum\limits_{k} {F_k(z) e^{- i k x/l}} \bigg\} \,,
\end{split}
\end{equation}
where $k = 2 \pi n/l$ with $n$ being a positive integer.  These series expansions can be done because the horizontal extent of the domain is finite, and they automatically satisfy the boundary conditions for the $x$-direction \eqref{qbdyN0}.  This allows the dimensionality of the problem to be reduced to one. Then \eqref{qeqnN0} and the remaining boundary conditions for the vertical direction \eqref{qbdyN0} become
\begin{align}
\frac{\partial^2 Q_k}{\partial z^2} - \kappa^2 Q_k =& - F_k, \label{Qeqn}
\end{align}
\begin{align}
\frac{\partial Q_k}{\partial z}  - \frac{N_0^2}{2g} Q_k \, \bigg|_{z=0} = \frac{\partial Q_k}{\partial z}  - \frac{N_0^2}{2g} Q_k \, \bigg|_{z=h} = 0, \label{Qbdy}
\end{align}
where $\kappa^2 = k^2 + N_0^4/4 g^2$. Solving for $Q_k$ for each mode $k$ and summing over all the modes gives us $q$ which will then give $p$, the pressure perturbation field. 

Equation \eqref{Qeqn} can be solved by taking a Green's function approach. This is as far as we can take the solution analytically, since the source term $F_k$ in \eqref{Qeqn} is given from laboratory data. The Green's function $G_k$ for this case satisfies
\begin{align}
\frac{\partial^2 G_k}{\partial z^2} - \kappa^2 G_k =& \delta(z-z'), \label{Geqn}
\end{align}
\begin{align}
\frac{\partial G_k}{\partial z}  - \frac{N_0^2}{2g} G_k \, \bigg|_{z=0} = \frac{\partial G_k}{\partial z}  - \frac{N_0^2}{2g} G_k \, \bigg|_{z=h} = 0. 
\label{Gbdy}
\end{align}
Considering  \eqref{Geqn} on each side of the jump, 
\begin{align}
\frac{\partial^2 G_k}{\partial z^2} - \kappa^2 G_k =& 0\,,
 \label{Geqn2}
\end{align}
gives a solution of  the form
\begin{align}
G_k(z,z') =
\begin{cases}
G_k^{z>z'}=A e^{\kappa z} + B  e^{-\kappa z}, & z > z' \\
G_k^{z<z'}=C e^{\kappa z} + D  e^{-\kappa z}, & z < z',
\end{cases}
\label{Green}
\end{align}
where the constants $A,B,C$,  and $D$ are determined  by the  following matching conditions:
\begin{align}
G_k^{z>z'}(z,z')\,\bigg|_{z=z'} = G_k^{z<z'}(z,z')\,\bigg|_{z=z'}\,, 
\label{Gk1} \\
\frac{\partial}{\partial z} G_k^{z>z'}(z,z') \, \bigg|_{z=z'} = 1 + \frac{\partial}{\partial z} G_k^{z<z'}(z,z') \, \bigg|_{z=z'}\,.
 \label{Gk2}
\end{align}
After applying the matching conditions \eqref{Gk1}, \eqref{Gk2}, and the boundary conditions \eqref{Gbdy}, the following Green's function \eqref{Green} for mode $k$  is obtained:
\begin{equation}
G_k(z,z') = 
 \frac{1}{\gamma} \left[ \kappa_+^2 \, e^{\kappa z_{+}} + 2 k^2 \cosh{(\kappa z_{-})} +\kappa_-^2\, e^{-\kappa  z_{+}} 
 \right]\,,
 \label{greenefinal}
 \end{equation}
where   $z_{+} = z + z'-h$, $z_{-} = |z - z'|-h$,   $\gamma = - 4 \kappa k^2 \sinh{\kappa h}$, and $\kappa_{\pm}=\kappa\pm {N_0^2}/({2g})$.

The solution is obtained by convolving  $G_k$  with $F_k$ (which is given in terms of the perturbation density $\rho$ from \eqref{f}) to find the $Q_k$ in \eqref{Qeqn}, which are the Fourier coefficients for $q$ in \eqref{qeqnN0}, which then can be transformed to find the pressure perturbation field $p$, 
\begin{equation}
\begin{split}
p(x,z) &= \mathrm{Re} \, \bigg\{ - \frac{2}{l} e^{ - N_0^2 z / 2g}   \\ 
&\phantom{=} \hspace{-1.5cm} \times \sum\limits_{k} { e^{- ikx} \int_{0}^{h}{ dz' \, G_k(z,z') \, \int_{0}^{l}{dx' \, f(x',z') \, e^{ikx'}} }} \bigg\}, \label{greenspressure}
\end{split}
\end{equation}
where $k = 2 \pi n / l$, with $n$ a positive integer, and  $f$, recall, is determined by $\rho$ according to  \eqref{f}. 


\section{Numerical simulations and laboratory experiments}
\label{sec:Methods}


To test our approach and to explore its robustness, we apply it to density perturbation data for both numerically simulated and experimentally measured internal wave beams. The numerical simulations are described in section~\ref{sec:simulations}, while the laboratory tank system and synthetic schlieren measurements are described in section~\ref{sec:expt}.  Comparison of the density perturbation fields from the simulations and the synthetic schlieren measurements is made in section~\ref{sec:sim_exp_comp} in order  to validate the application of our method to laboratory data.


\subsection{Navier-Stokes numerical simulations} 
\label{sec:simulations}


Our direct numerical simulations of the Navier-Stokes equations  yield density, velocity, and pressure perturbation fields for a system with a driven internal wave beam.   The  energy flux computed from these fields will be compared to the values obtained by the approach that uses only density perturbation data, as described in section~\ref{sec:analytical}.  The simulations use the CDP-2.4 code, which solves the Navier-Stokes equations in the Boussinesq approximation~\citep{ham04}.  This finite-volume based solver implements a fractional-step time-marching scheme, with subgrid modeling deactivated.  The code has been validated in previous laboratory and computational studies of internal waves~\citep{king09,lee14,dettner13,zhang14,paoletti14}.

\begin{figure*}
\includegraphics[width=\textwidth]{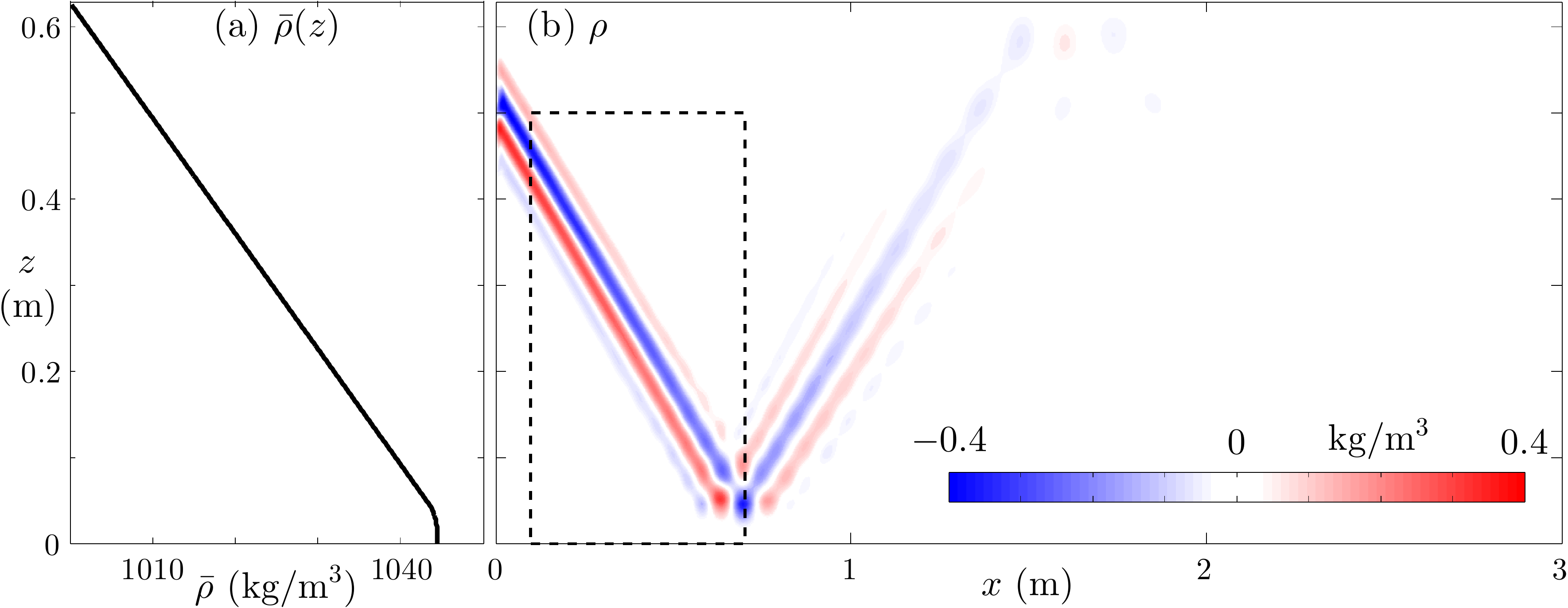}
\caption{ (a) The density profile used for both the experiment and simulations; the buoyancy frequency is constant, $N=0.8533$ rad/s, except $N=0$ in a layer about 0.04 m thick at the bottom. (b) Simulation results for the density perturbation field from the internal wave generated in the upper left corner.  The simulation domain is a rectangular box that extends from -3 m to +3 m, while the laboratory schlieren measurements are made in a region that corresponds to the box bordered by dashed lines.  In this snapshot, made at an instant after 11.75 periods of forcing, the internal wave beam has reached a steady state in the region of the schlieren measurements, but the flow is still evolving in the region to the right of the dashed box.} \label{fig1}
\end{figure*}

The simulations are conducted in a two-dimensional domain with $x \in [-3.0, 3.0]$ m and $z \in [0, 0.63]$ m.  Domain dimensions and parameters for the simulation are selected for comparison with the experiment discussed in section~\ref{sec:expt}. The simulation solves the following for the total density $\rho_T$, pressure $p_T$, and velocity $\bm{u}_T$:
\begin{align}
\frac{\partial \bm{u}_T}{\partial t} + \bm{u}_T\cdot\nabla\bm{u}_T &= -\frac{1}{\rho_{00}}\nabla p_T + \nu\nabla^2 \bm{u}_T - \frac{g \rho_T}{\rho_{{00}}}\bm{\hat{z}}, \\
\frac{\partial \rho_T}{\partial t} + \bm{u}_T\cdot\nabla\rho_T &= \kappa \nabla^2 \rho_T, \nabla \cdot \bm{u}_T = 0,
\end{align}
where $\rho_{00}=1000 $ kg/m$^3$ (density of water), $\nu=10^{-6}$ m$^2$/s (kinematic viscosity of water at $20^o$C), and $\kappa=2\times10^{-9}$ m$^2$/s (the diffusivity of NaCl in water).   Initially the system is stationary with a linear density stratification with buoyancy frequency $N=0.8533$~rad/s, except in the bottom $0.04$ m where the density is constant (figure 1($\it a$)).  The boundary conditions are free-slip at the top and no-slip at the bottom. The left and right boundaries are periodic with Rayleigh damping, proportional to the velocity, implemented within 0.5~m of the left and right ends of the domain,  preventing any advection through the boundary.  

An internal wave beam is produced using a momentum source in $x \in [-0.01,0.01]$~m and $z \in [0.43,0.5825]$~m that imposes the velocity
\begin{align}
\bm{u}_T &= \omega A(z) \sin(\omega t - k_z z)\bm{\hat{x}} \label{uexp},
\end{align}
with an amplitude profile given by
\begin{align}
A(z) &= \exp(-(z-0.50625)^2/0.22), \label{Az}
\end{align}
where the lengths are in meters, and $k_z=82.45$ m$^{-1}$.  A time step $\delta t=0.0025$~s (5200 steps per period) is sufficient for temporal convergence. Spatial convergence is obtained using a structured mesh with resolution $\delta x\approx 10^{-7}$ m near the boundaries,  $\delta x \approx 10^{-4}$ m within the internal wave beam, and $\delta x \approx 10^{-2}$ m away from the active region.  Changes in the velocity field are less than $1\%$ when spatial and temporal resolutions are doubled.

A snapshot of the density perturbation field from the simulation is presented in figure~\ref{fig1}($\it b$).  Only the right half of the domain is shown because the system is symmetric about $x=0$ m.  The internal wave beam is produced at $x$ = 0 m at a height of about $z$ = 0.5 m, and the reflection of the beam occurs at $(x,z)$ = $(0.7,0.04)$ m.  The constant density layer in the bottom $0.04$ m does not propagate waves because the forcing frequency is higher than the local buoyancy frequency.  This snapshot is taken after 11.75 periods of forcing, which is sufficiently long for the beam to reach the bottom of the domain but not yet reach a steady state.
 

\subsection{Experimental techniques}
\label{sec:expt}


The intended application of the approach is for observed data either in the ocean or in a tank experiment.  A tank-based experiment analogous to the simulation is performed where synthetic schlieren measurements are made to obtain the instantaneous density perturbation field.  

\begin{figure*}
\includegraphics[width=\textwidth]{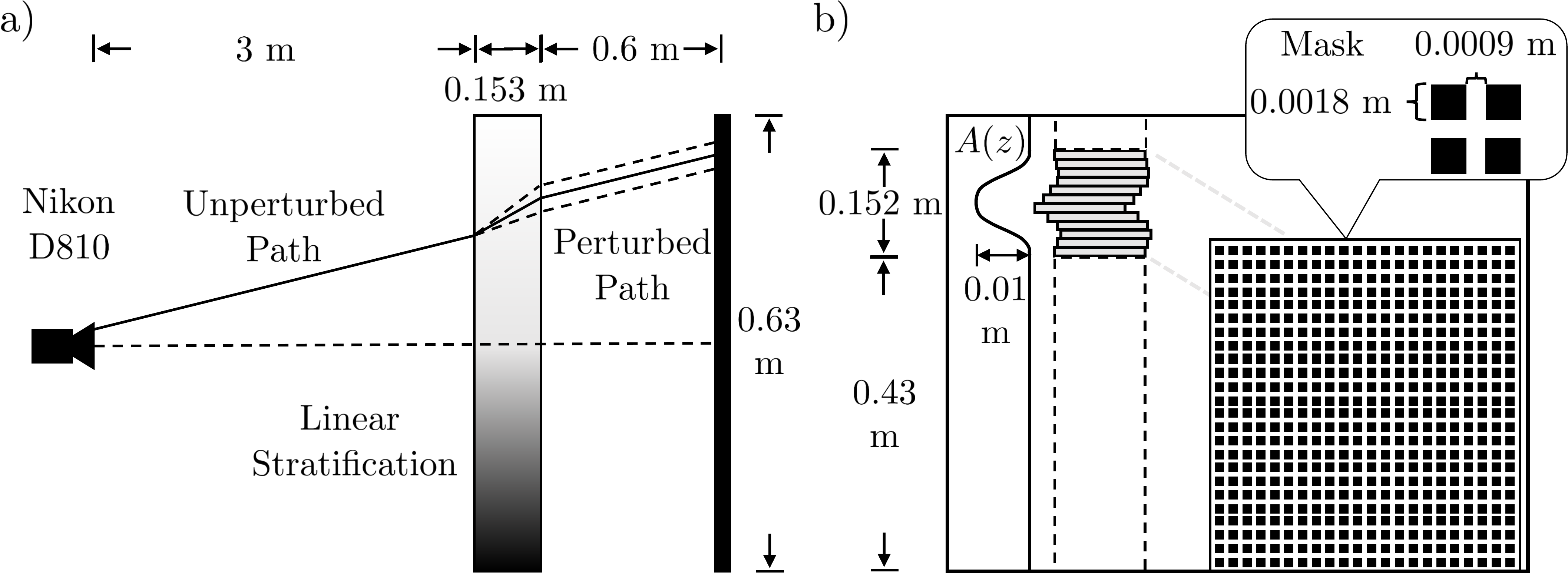}
\caption{(a)  A sketch of the experimental system.  The camera observes, through the stratified fluid, a white screen located 0.6 m beyond the tank.  The screen is covered by a mask (shown in (b)), and is back-lit by a panel of LEDs.  Density perturbations caused by the internal wave beam change the fluid index of refraction, causing the mask to appear to move, and digital movies record this motion.  (b)  The internal wave generator has 12 plates that are driven by a camshaft, and each cam is an eccentric disk on a hexagonal rod that is rotated by a stepper motor.  The disk eccentricity, $A(z)$, is a Gaussian profile.  The mask covering the LED panel is a rectangular array of black squares, each 0.0018 m $\times$ 0.0018~m with 0.0009~m gaps in between.}\label{fig2}
\end{figure*}

The laboratory system for determining the density perturbation field by the synthetic schlieren method is diagrammed in figure~\ref{fig2}($\it a$): a density-stratified fluid is contained in a lucite tank that has interior dimensions of 4 m $\times$ 0.7 m $\times$ 0.15 m, and the apparatus for generating internal waves (figure~\ref{fig2}($\it b$)) is 3 m from the end of the tank.  The tank is filled slowly from the bottom, using the generalized double-bucket procedure of \citet{hill02}, which uses two fluid reservoirs, one with pure water and the other with saturated salt water, to produce the desired fluid density profile. In our tank, the density increases linearly from 1000 kg/m$^3$ (pure water) at the top to a density of 1045 kg/m$^3$ (salt solution) at a height just 0.04 m above the bottom; below 0.04 m the density is approximately constant (see figure~\ref{fig1}($\it a$)).  The constant density layer is added to lift the fluid away from optical distortions at the bottom of the tank.  To measure the stratification, fluid samples are withdrawn from the tank at various heights and their densities are measured with an Anton-Parr density meter.

An internal wave beam is generated with a camshaft-driven wavemaker based on the design of \citet{mercier10} (see figure~\ref{fig2}($\it b$)).  A rotating camshaft drives a stack of 12 delrin plastic plates (cams) to produce a  velocity profile approximating the one used in the simulations.  The cams are $0.0762$ m diameter circular disks that are offset from their centers by distances prescribed by equation~\eqref{Az}.  The hexagon drive shaft gives a phase difference of $\pi/3$ between consecutive disks.  The wavemaker is driven at $(2\pi)/13$ rad/sec, which yields a beam with an angle of $\theta=34.5^{o}$ with respect to the horizontal, based on the dispersion relation $\sin\theta=\omega/N$.  

The density perturbation field resulting from the two-dimensional internal wave beam is observed using the synthetic schlieren method, which uses the linear relationship between the local density gradient and the index of refraction of the density-stratified fluid \citep{sutherland99, dalziel00}.  The distorted images of the mask's square grid pattern  (cf. figure~\ref{fig2}) are recorded with a camera on the opposite side, as in~\citet{sutherland14}. Calculation of the corresponding density perturbation field through integration, however, has proven to be challenging because the time-dependent image must have a large signal-to-noise ratio in order to obtain accurate density perturbation fields to implement the method described in section~\ref{sec:Theory}.  As a result of this challenge, only a few investigations have actually calculated density perturbation fields from schlieren measurements~\cite{dalziel07, yick07, hazewinkel11, jia14}.  
  
To allow us to accurately integrate the density-perturbation field, we achieve a large signal-to-noise ratio using a Nikon D810 camera with 7360 $\times$ 4912 pixels to image the pattern of black squares of the mask (see figure~\ref{fig2}($\it b$)).  The camera is placed 3~m in front of the tank.  The D810 camera has focus and mirror locks that reduce camera and focus jitter during closure of the mechanical shutter.  The camera images a 0.86 m $\times$ 0.51 m region that starts 0.1 m to the right of the wavemaker and extends upward from the bottom of the tank.  Images are taken at a frequency of 1 Hz, which corresponds to 13 images per wave period.  There are 10 pixels across each black square in the image; in the quiescent system the image of a black square moves less than 0.1 pixel due to thermal variations and camera shake. In the most intense part of the internal wave beam the black squares are displaced typically by 6 pixels.

The positions of the individual black squares in the images are determined with subpixel accuracy using a particle tracking code that identifies centers of squares by a least-squares method~\citep{shattuck2015}. To create the displacement values, reference positions of the squares are determined from a sequence of images obtained before the wavemaker is turned on. Then the displacement field of the squares is computed from the images in the digital movie, and the displacements are used to calculate perturbations of $\nabla \rho$.  Through application of a partial-differential-equation solver that eliminates the rotational noise in the measurements, the density-gradient perturbations are used to calculate a density perturbation field~\citep{hazewinkel11}.  While we performed the calculation independently, the density perturbation field can be computed from schlieren data using the software package DigiFlow~\citep{digiflow}. 


\subsection{Comparison between simulation and experiment}\label{sec:sim_exp_comp}


\begin{figure*}
\centering
\includegraphics[width=\textwidth]{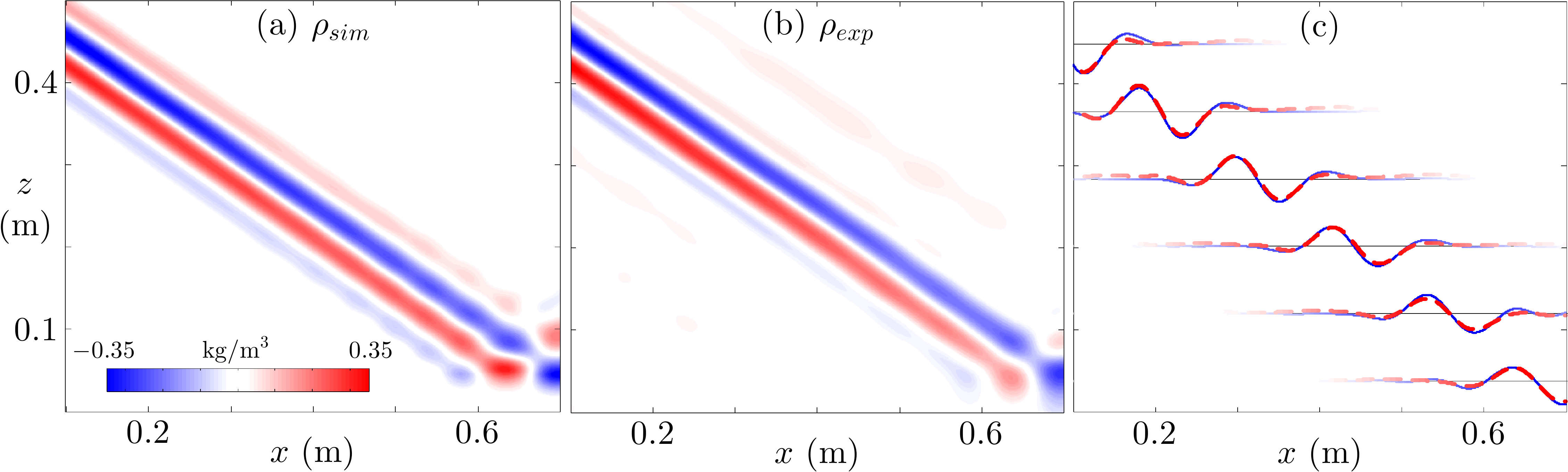}
\caption{The instantaneous density perturbation field from (a) simulation  and (b) experiment. (c) The synthetic schlieren density perturbation measurements (red dashed) agree well with the numerical simulation results (blue solid) at different heights in the tank.  The horizontal black lines correspond to zero perturbation.  The maximum amplitude of the perturbation is 0.36 kg/m$^3$.}
\label{fig3}
\end{figure*}

Care was taken to match the conditions of the experiment and simulation, but there are differences, particularly in the layers of nearly constant density at the top  and bottom of the laboratory tank. The laboratory wavemaker forcing profile modeled by equation~\eqref{Az} was fit to the eccentricity profile used in the experiments, but the match was not perfect. However, the frequencies were accurately matched.  Another minor difference between the simulation and experiment is that the free surface in the experiment falls and rises about 10$^{-7}$~m, while the simulation compensates for the small periodic volume flux with a background flow that is at least five orders of magnitude smaller than the velocities in the beam. Finally, our comparisons between the simulation and experiment are made at an early enough time that the internal wave beam has not reflected off the far end of the tank.

The simulated density perturbation field matches well with the laboratory schlieren data obtained in the region corresponding to the black dashed box of figure~\ref{fig1}, as can be seen by comparing figures~\ref{fig3}$(\it{a})$ and $(\it{b})$.  The amplitude of the experimentally measured density perturbation is $2\%$ smaller than in the simulation.  The experimental internal wave beam has a narrower band of large density perturbation, which is perhaps due to weaker realized forcing by the top and bottom plates of the wavemaker.  Finally, the density perturbation below the reflection region differs from the experimental internal wave beam, which penetrates further into the bottom near-constant density layer.

The simulated and experimental density perturbation profiles at six heights are compared in figure~\ref{fig3}$(\it{c})$.  The rms difference (relative to the beam amplitude) between the simulated and measured density perturbation fields within the beam is about $9\%$, except near the bottom of the tank where the difference rises to as much as $30\%$. The large error in the constant density layer at the bottom boundary arises because, as aforementioned, the simulation density profile there does not precisely match the density profile in the tank.  
 

\section{Results} \label{sec:Results}


Given the density perturbation fields from section~\ref{sec:simulations}, we obtain the instantaneous velocity, pressure, and energy flux using our method, and compare them to the simulated results in section~\ref{subsec:full}.  This verification of the method presented in section~\ref{sec:Theory} uses the entire simulation domain, which satisfies the boundary conditions in equations \eqref{pbdy1} and \eqref{pbdy2}.  Then section \ref{subsec:Labdata} applies the method to laboratory schlieren measurements of the density perturbation field.  These calculations are made in a subdomain of the simulations, but we show that with appropriate buffering of the laboratory data the results for the energy flux determined by the method agree well with direct Navier-Stokes simulations.


\subsection{Verification of the method by comparison with direct numerical simulations} \label{subsec:full}

\begin{figure*}
\centering
\includegraphics[width=\textwidth]{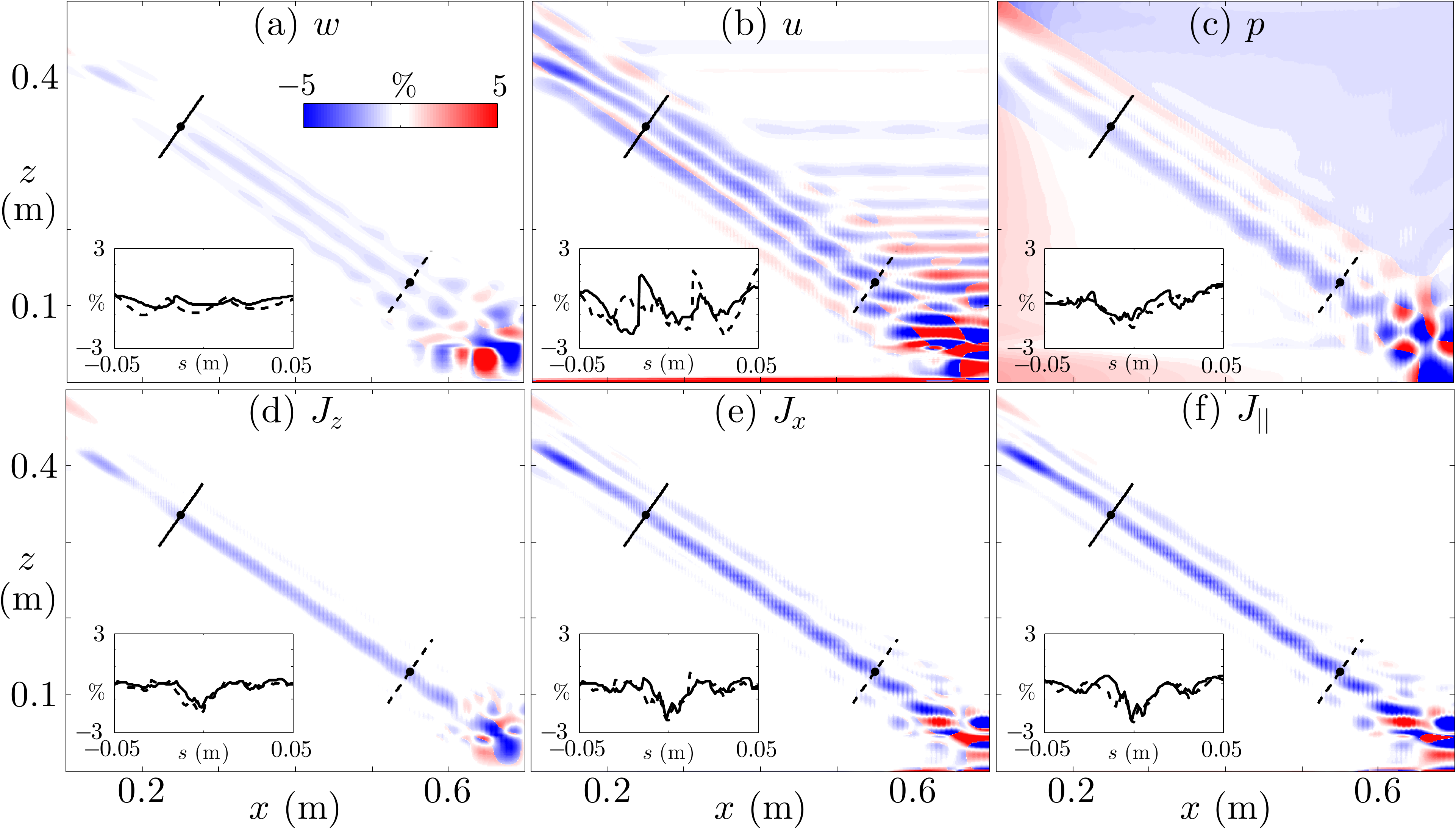}
\caption{The percent difference (relative to the peak amplitude) of the instantaneous fields calculated solely from the simulation density perturbation field compared with the direct Navier-Stokes simulation values for $w$ (a), $u$ (b), $p$ (c), ${J_z}$ (d), ${J_x}$ (e), and the energy flux component parallel to the beam,  ${J_{\parallel}}$ (f). The insets show profiles in the beam towards the top-left (solid) and bottom-right (dashed) corners of the domain. Excluding the reflection region where the buoyancy frequency deviates, the difference is within 3$\%$ for all quantities.} 
\label{fig4}
\end{figure*}

The Green's function method for determining the instantaneous velocity perturbations, pressure perturbations, and energy flux from density perturbation data for internal waves is verified by comparison with results from the numerical simulations. As figure~\ref{fig4} shows, the fields $w$, $u$, and $p$  calculated solely from simulation density perturbation data agree with the direct simulation values typically to within a few percent, and the results for the energy flux $\bm{J}$ agree with the simulations to within $1\%$ throughout most of the domain, except in the thin constant density layer near the bottom. There the buoyancy frequency profile deviates from the uniform value of the rest of the domain. Note that all percent differences are relative to the peak amplitude.  The analysis is performed on the internal wave field in the entire domain in figure~\ref{fig1} to satisfy the boundary conditions \eqref{pbdy1} and \eqref{pbdy2}.

The vertical velocity component $w$  in figure~\ref{fig4}$(\it{a})$ is straightforwardly obtained from the time derivative of the density perturbation field \eqref{w}.  Throughout the domain the results closely match, and across the beam the rms percent difference (normalized by the peak amplitude) between the density-calculated and simulated values is $0.8\%$. The largest errors occur where the wave beam is generated and in the region where the beam reflects from the thin constant density layer at the bottom (cf. figure~\ref{fig1}($\it b$)). In the latter region the percent difference is as high as $11\%$.

The horizontal velocity component $u$ is calculated by integrating the incompressibility condition with the previously calculated $w$ of \eqref{u}.  Taking initial integration points where the velocity is known to be zero or small, the normalized rms difference between $u$  from the density-calculated method and from direct simulations is $2.2\%$ across the internal wave beam. The amplitude-normalized percent difference is less than $2\%$ throughout the beam but reaches errors as large as $26\%$ at the constant density layer interface.  However, because we assume the starting point has zero velocity, Any error in our assumption that the starting point has zero velocity will propagate across the horizontal slice, as is evident to the right of the beam. 

The first step in determining the pressure perturbation field from the density perturbation field is the calculation of the Fourier coefficients of $f(x,z)$ [(equation (3.20)] for each horizontal slice of the domain (cf. equation \eqref{greenspressure}).  We find that 300 modes are sufficient for convergence for the high resolution simulation data with a small beam width relative to the domain width.  The Fourier coefficients are then used in the Green's function calculation to obtain the pressure perturbation field $p$.  The normalized rms difference between this calculated $p$ and the value of $p$ direct from the simulations is $3\%$ in the beam (figure~\ref{fig4}$(\it{c})$). Again the largest errors ($11\%$) are in the regions of wave beam generation and reflection.

\begin{figure*}\centering
\includegraphics[width=\textwidth]{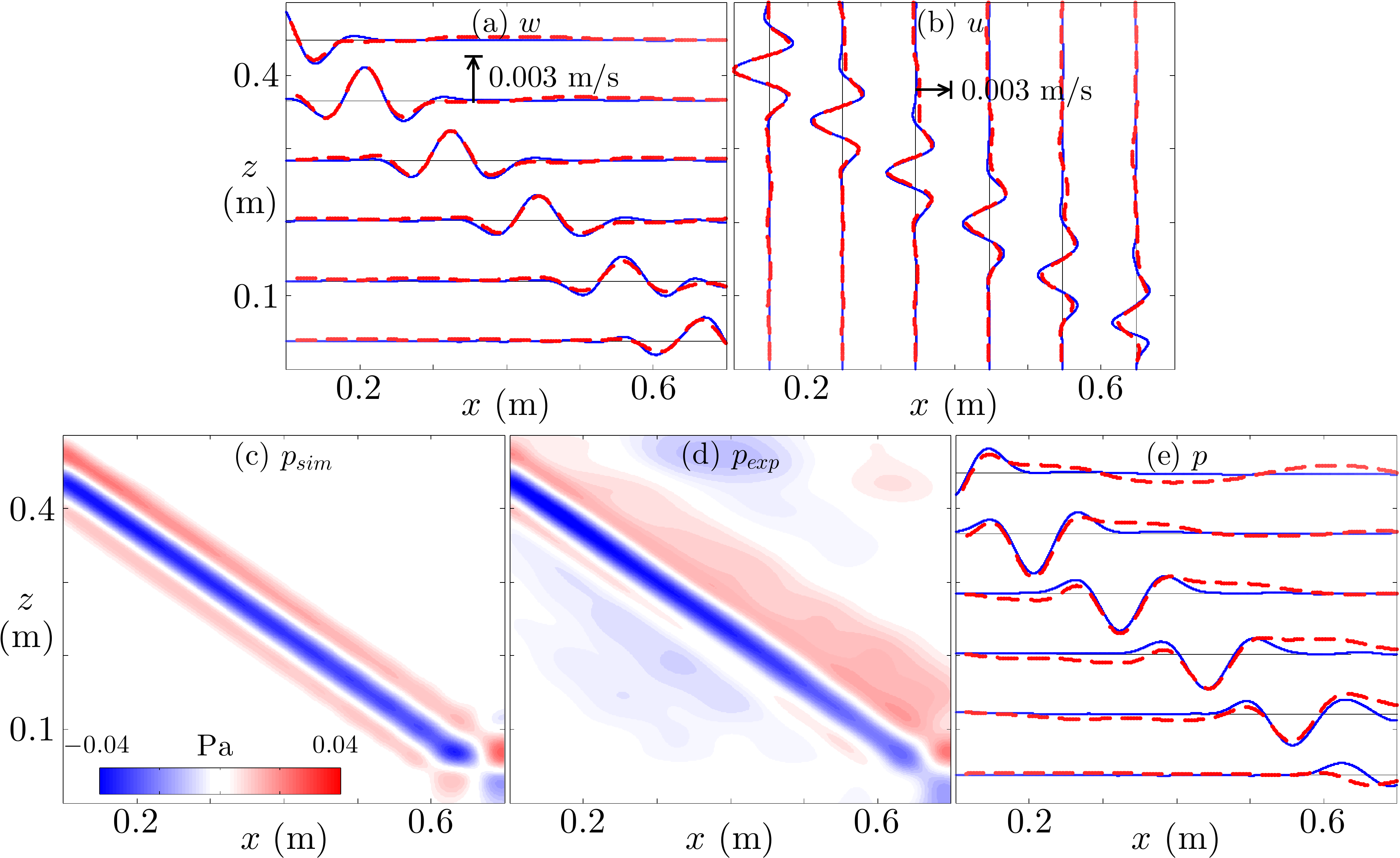}
\caption{Experimental (red dashed) and simulation (blue solid) results at different heights are compared for $w$ (a) and $u$ (b). (c) The pressure perturbation field from the simulation. (d) $p$ from the Green's function method applied to laboratory data. (e) A comparison of the results in (c) and (d) at different heights. }
\label{fig5}
\end{figure*}

Finally, the energy flux is obtained by multiplying the calculated velocity and pressure perturbation fields.  Figures~\ref{fig4}$(\it{d})$ and $(\it{e})$ compare ${J_z}$ and ${J_x}$ obtained from the density perturbation field with the direct numerical simulations, respectively. The normalized rms difference in the vertical energy flux in the internal wave beam is $0.8\%$, which matches the precision of the vertical velocity calculation.  The maximum difference in the flux magnitude occurs in the reflection region and is $4.5\%$ (cf. figure~\ref{fig4}$(\it{f})$), which is lower than the individual components because the overestimate of the calculated vertical velocity is partially compensated by an underestimate of the pressure. Throughout most of the beam the normalized percent difference between our method and the direct Navier-Stokes simulation result for the energy flux is less than $1.0\%$.  Because the calculation of the velocity and pressure tend to underestimate the actual values, the energy flux is also underestimated.


\subsection{Application of the Method to Laboratory Data} \label{subsec:Labdata}


Having verified the method in the previous subsection, we now apply it to the experimental data presented in section~\ref{sec:sim_exp_comp}.  The data is obtained in the portion of the domain within the black dashed box in figure~\ref{fig1}, but this subdomain does not satisfy the boundary conditions taken for the method.  However, in appendix~\ref{sec:cropped} we present a procedure that  accommodates data sets for subdomains that do not strictly satisfy the boundary conditions. For better comparisons between the simulated and experimental results, the simulation data in this subsection uses a lower data resolution, which is identical to that of the experiment. As mentioned in section~\ref{sec:sim_exp_comp}, the measured and simulated density perturbation fields are not identical, but closely represent the same instant allowing the use of the simulated results for comparison of the velocity perturbation, pressure perturbation, and energy flux fields.

\begin{figure*}
\centering
\includegraphics[width=\textwidth]{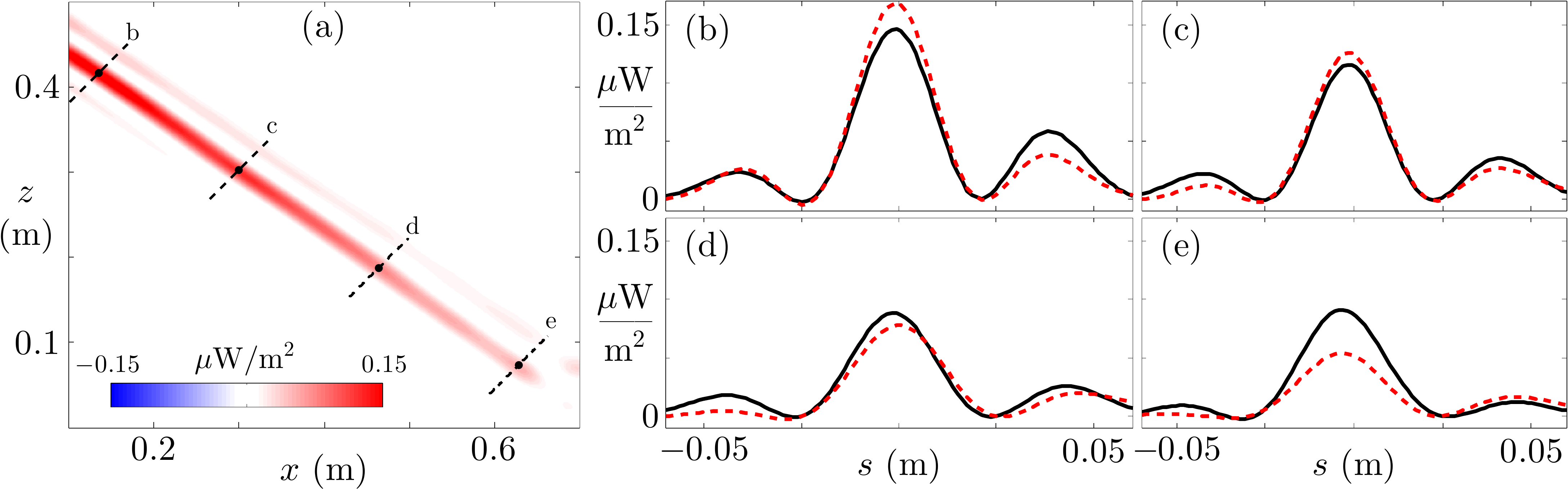}
\caption{(a) The energy flux in the direction of the internal wave beam, obtained using the Green's function method on the experimental density perturbation data. (b)-(e) The energy flux in cross sections of the beam, computed from the Navier-Stokes simulations (black solid) and from the Green's function method on the laboratory measurements (red dashed). The agreement is very good in (c) and (d), but less so in regions where the simulations have less accurate representations of the laboratory system, that is, near the internal wave source (b) and near the unstratified the thin bottom layer (e).}   
\label{fig6}
\end{figure*}

The velocity components from the simulations and laboratory measurements are compared in figures~\ref{fig5}$(\it{a})$ and $(\it{b})$.  The camera was limited to 13 frames per period, but despite this large time step the results calculated from the lower-resolution simulation data for the time derivative of the density perturbation differ from the high-resolution results presented in section~\ref{subsec:full} by less than $1\%$.  The vertical velocity profiles from the simulation and experiment in figure~\ref{fig5}$(\it{a})$ have an average normalized rms difference of $8.1\%$ in the beam.  The horizontal velocity profiles in figure~\ref{fig5}$(\it{b})$ have similar average normalized rms differences, $8.4\%$.  The largest error, as much as $30\%$, occurs in the reflection region where the simulation and laboratory density stratification profiles differ. 

Outside the beam the velocity field calculated from the experimental density perturbation field agrees well with the values direct from the simulations. However, outside the beam the pressure perturbation field $p$ found by applying the Green's function method to the experimental data does not agree as well with the corresponding values from the numerical simulation, as figures~\ref{fig5}$(\it{c})$ and $(\it{d})$ show. The differences between $p$ from the simulation and the experiment result primarily from the differences in the lower mode Fourier components, because of error at larger length scales in the experimental density perturbation data (not shown).  The resultant difference is evident in the plots of $p$ at different heights in figure~\ref{fig5}$(\it{e})$.  The average normalized rms difference in $p$ in the beam and for the full domain are comparable, $15.1\%$ and $14.0\%$, respectively.

Despite the differences in $p$ direct from the simulation and the Green's function calculation of the laboratory data, the energy flux obtained by the Green's function method differs from the simulation typically by only $6\%$ (rms difference normalized by the flux amplitude), as figure~\ref{fig6} shows. The  Green's function result for the flux outside of the beam does not have the artifacts present in the pressure field, because in those regions the velocity is close to zero. The agreement is not as good at the upper left (cf. figure~\ref{fig6}(b)), where the laboratory internal wave generator is represented by an approximate model form in the Navier-Stokes simulations, and at the lower right where the beam reflects from a thin unstratified bottom layer, which is also only modeled approximately in the Navier-Stokes simulations (cf. figure~\ref{fig6}(e)).


\section{Conclusions} \label{sec:Conclusions}


We have presented a Green's function method for calculating the instantaneous energy flux field $\bm{J} = p \bm{u}$ solely from the density perturbation field for linear internal waves in a density-stratified fluid with a uniform buoyancy frequency $N$. $\bm{J}$ is obtained from the density perturbation field through separate computations of $p$, $u$, and $w$:  $p$ using the Green's expression of (\eqref{greenspressure}), $w$ from the continuity equation \eqref{w}, and $u$ from incompressibility and knowledge of $w$ from the previous calculation. The method was verified using numerical Navier-Stokes simulations of our laboratory experiment on internal waves generated in a tank with a linearly stratified density fluid.  In most of the domain, $w$, $u$, $p$, and $\bm{J}$ calculated using the Green's function method solely from the density perturbation field from a Navier-Stokes simulation agree within a few percent with results obtained directly from the simulation. However, in regions near the wave generator and the unstratified bottom fluid layer, the results obtained directly from the simulations and from the Green's function method differ by as much as $5\%$.

The Green's function method was then applied to laboratory schlieren data. In order to match the boundary conditions  in the derivation, \eqref{pbdy1} and \eqref{pbdy2}, we used data buffers described in appendix \ref{sec:cropped} because the observational window for the schlieren measurements did not span the entire tank. The density perturbation field determined from the schlieren data differs from the numerical simulation by about $11\%$, but a counterbalance of errors in the velocity and pressure fields led to energy flux values from the experiment that agree with the numerical simulations to within $6\%$.

The Green's function method developed here was applied to internal waves in a linearly stratified fluid (uniform buoyancy frequency) and an analytic solution was found.  However, the theory in  section~\ref{sec:genTheory} applies to {\it any} stratification. Systems with nonlinear stratifications can be analyzed numerically with \eqref{J2}, and for some buoyancy frequency profiles $N(z)$ analytic solutions may also be possible.

While the method was applied here to a single internal wave beam, it also was found to work for a  wave field where a parametric subharmonic instability produced wave energy at two new wavenumbers and frequencies; this would be difficult to treat by time-averaged methods.   The present method can also be extended to systems with a background velocity, such as tidal flow.  Another extension would be to large-amplitude propagating internal waves such as internal solitary waves. Yet another interesting extension would be to weakly three-dimensional density perturbation fields, such as those that occur near ocean ridges and in coastal waters. 

To aid in the application of this method, a Matlab GUI has been developed, as described in appendix~\ref{sec:gui}.  Implementation of the GUI requires the density perturbation field, the coordinates of the data, the time step size, and the buoyancy frequency (which is assumed to be constant).  If a data set does not satisfy the boundary conditions assumed in our analysis, the GUI can implement the buffering technique used on our data and discussed in appendix~\ref{sec:cropped}.   The GUI  includes an operations manual and also a tutorial which recalculates the numerical results from section~\ref{subsec:Labdata}.

\section*{Acknowledgments} 

We thank Likun Zhang and Robert Moser for helpful discussions. The computations were done at the Texas Advanced Computing Center. MRA and HLS were supported by the Office of Naval Research MURI Grant N000141110701, while FML and PJM were supported by the U.S. Department of Energy, Office of Science, Office of Fusion Energy Sciences, under Award Number DE-FG02-04ER-54742.

\appendix


\section{Cropped domains and buffering}
\label{sec:cropped}


Density perturbation data from synthetic schlieren measurements are often from regions that do not contain the boundaries of the fluid system, and the boundary conditions \eqref{pbdy1} and \eqref{pbdy2} used to find the pressure perturbations are in general not satisfied on the boundaries of a `cropped'  measurement window. Cropping affects the calculation of pressure but not the  calculation of the vertical velocity field \eqref{w}, and as long as there is a point in the domain where the horizontal velocity is zero, the calculation for the horizontal velocity field \eqref{u} is unaffected as well. In this section, we use simulation data that have been cropped to test the effects on computations of the pressure and energy flux, and we present a procedure to minimize its impact.

\begin{figure}\centering
\includegraphics[height=.3\textheight]{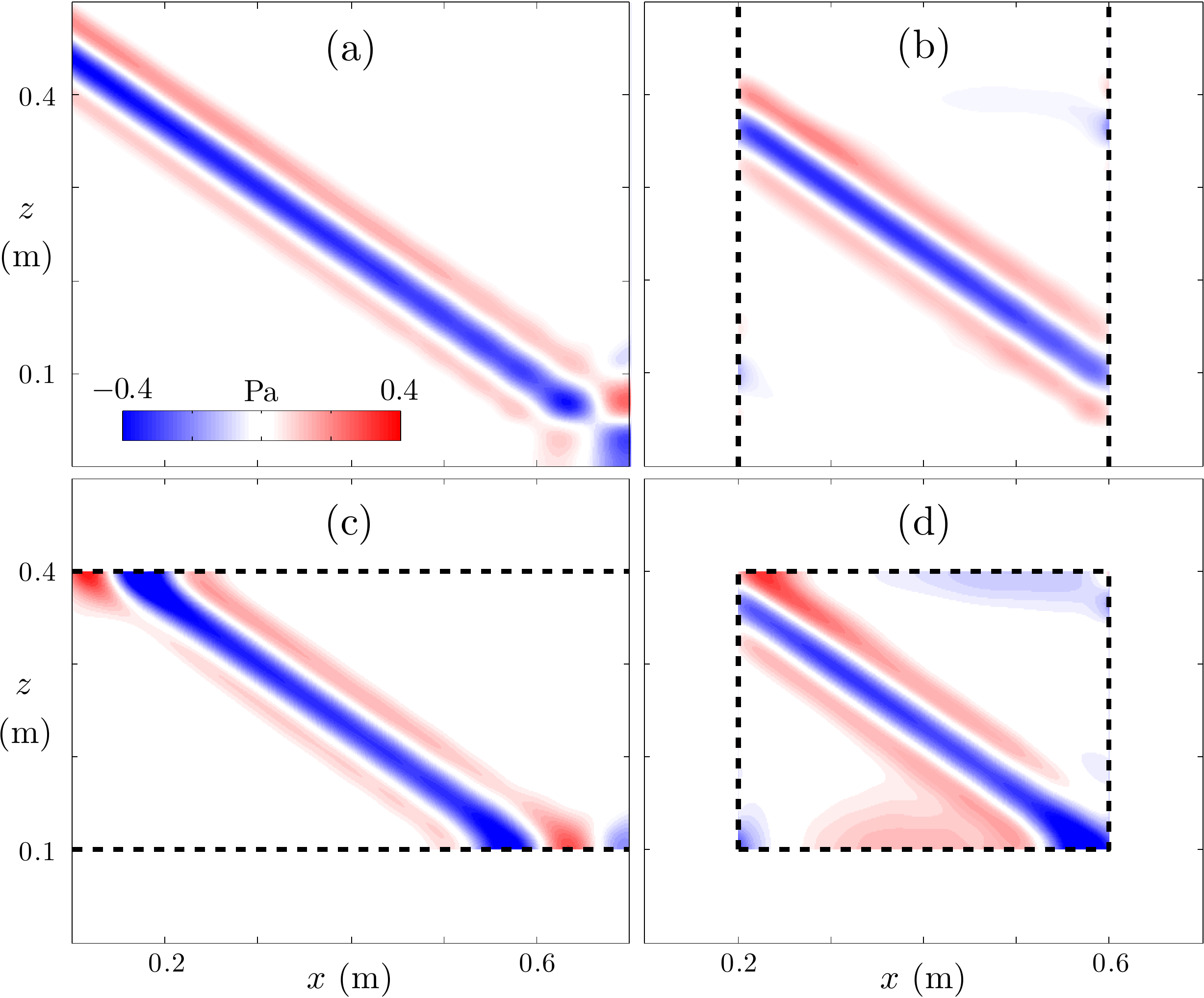}
\caption{Calculated pressure perturbation fields for the whole fluid domain (a) and for domains that have been cropped on the sides (b), the top and bottom (c), and both (d). The artifacts near the edges of the cropped domains (b), (c), and (d) arise from the violation of the boundary conditions \eqref{pbdy1} and \eqref{pbdy2}.}
\label{figA1}
\end{figure}

The Fourier series expansion in \eqref{Qeqn} reduces the dimensionality of the problem while respecting the boundary condition \eqref{pbdy1}.  Cropping the left and right sides of the domain in a way that results in the beam passing through the side boundaries will in general violate the periodic boundary condition and introduce a step discontinuity. Because the pressure perturbation is calculated as a Fourier series in the horizontal direction \eqref{greenspressure}, this cropping introduces Gibbs-phenomenon-like edge artifacts in the solution on the left and right boundaries. For reference, we show in figure~\ref{figA1}(a) the simulated pressure perturbation field in the domain used in the main body of the paper, and the impact of cropping the sides is shown in figure \ref{figA1}(b).  The edge artifacts resulting from the cropping can be present at the opposite end of the domain from where the beam penetrates, but the cropping does not significantly change the pressure field in the middle of the domain.

The boundary conditions at the top and bottom \eqref{pbdy2} are physically more important than those at the sides \eqref{pbdy1} because a no-flux condition is applied at the top and bottom for the Green's function \eqref{Gbdy}.  If the beam passes through the top and/or bottom boundary, then the no-flux condition is violated and error is introduced in the Green's function. Figure \ref{figA1}(c) shows that the resulting errors can be significant near the top and bottom boundaries, but again in the middle region the solution is quite good.
When the data are cropped in both directions, the errors from both the side and top-bottom cropping are present as one might expect, as shown in figure \ref{figA1}(d).

\begin{figure*}
\includegraphics[width=\textwidth]{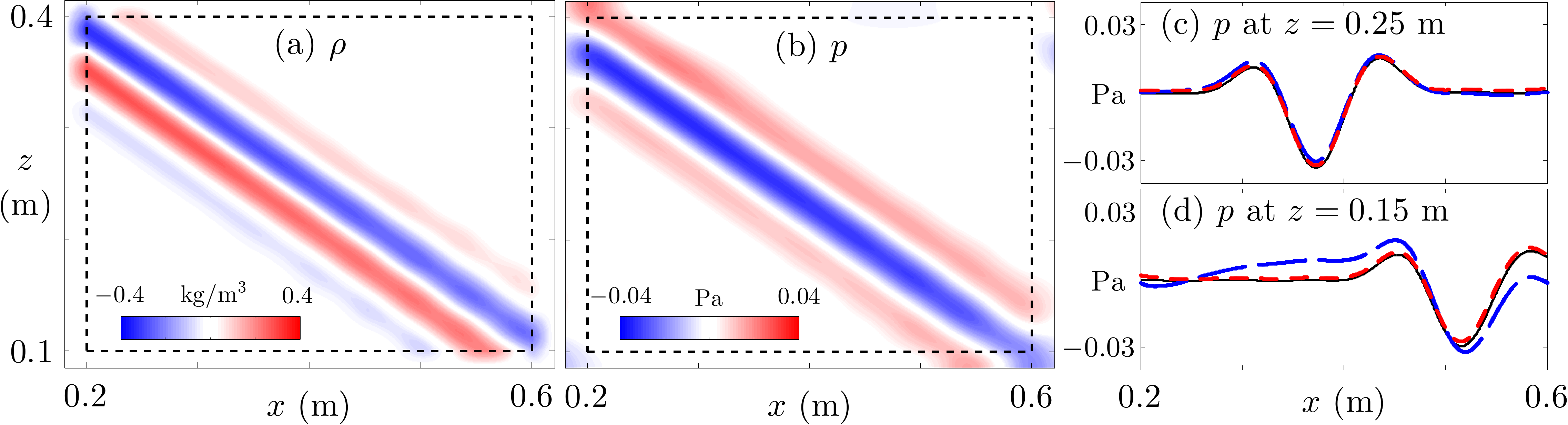}
\caption{(a) The density perturbation field calculated from the original data inside the black dashed box plus a 5\% buffer area.  (b) Pressure perturbation field (in Pascals, $Pa$) obtained from the buffered density perturbation data. This is much better than the un-buffered calculation from figure \ref{figA1}(d). Comparison of the simulated pressure perturbation (black solid) at heights $z=0.25$ m (c) and $z=0.15$ m (d) with the pressure perturbation calculated by the Green's function method both with a buffer (red dashed narrow) and without (blue dashed wide).} 
\label{figA2}
\end{figure*}

To minimize errors caused by cropping we introduce a method of buffering the data. This buffering is applied only to the pressure calculation as the velocity calculations do not depend on the boundary conditions.
Figure~\ref{figA2}$({\it a})$ shows an example of buffering the density perturbation field used to calculate the pressure perturbation in the cropped domain of figure~\ref{figA1}$({\it d})$.  The original domain inside the black dashed box is extended by $5\%$ in all directions.  The jump in density perturbation is removed by applying a smoothing filter on the new density perturbation field. In this smoothing process the density perturbation at the boundaries of the new domain is held at zero, and the values in the old domain are diffused into the expanded domain.  After applying the smoothing, the original density perturbation is reapplied, resulting in a density perturbation field that is unchanged within the original domain. The original density perturbation smoothly transitions to zeros along the edges, as shown in figure~\ref{figA2}~$({\it a})$.

The pressure perturbation calculation can then be applied to the extended domain and the results with a 5\% buffer region are shown for the density perturbation field in figure~\ref{figA2}$({\it a})$, and for the pressure perturbation field in figure~\ref{figA2}$({\it b})$.  For this small buffer size there are still some erroneous signatures in the top right and bottom left of the domain that are similar to the results from cropping the sides of the domain, but these errors are much smaller and are mostly contained in the buffer region. The addition of the buffer significantly reduces the error in the pressure perturbation calculation throughout the original domain.  Figure~\ref{figA2}$({\it c})$ shows that the results in the middle of the domain  are essentially the same with and without a buffer, but near the boundaries the benefit of the buffer is significant, as figure~\ref{figA2}$({\it d})$ shows. The normalized rms difference relative to the direct simulation results for the pressure perturbation calculation without the use of a buffer over the entire domain is $17\%$, while the addition of a 5\% buffer around the whole domain reduces the normalized rms error to $5\%$. Going further with a 20\% buffer reduces the error to $3\%$, which is comparable to the precision found in the verification (section~\ref{subsec:full}).  

Buffering the data domain seems to bring subtly different beneficial effects for the horizontal and vertical directions. The main benefit of buffering the left and right sides of the domain seems to be the removal of the step discontinuities at those boundaries. Since the original density perturbation source is Fourier expanded in the horizontal direction, the solution for the pressure perturbation is a Fourier series of Green's functions $G_k$ and their corresponding Fourier coefficient fields $F_k$. By removing the step discontinuities in the density perturbation field, the Gibbs-phenomenon-like edge effects in the series solution for the pressure perturbation is significantly reduced. However, this means that excessive buffering in the horizontal direction (approaching the horizontal length-scale of the beam) can artificially introduce lower $k$ modes and produce errors.

The main benefits of buffering the top and bottom of the domain seem to be to push the no-flux boundary away from the original boundary, and to produce an extension of the beam that somewhat mimics the original density perturbation. Pushing the no-flux boundary away makes the Green's function behavior more appropriate for a beam that does not reflect at the boundary. The extension of the beam in the buffer region provides an approximate source that, combined with the aforementioned improved Green's function, produces a better result for the pressure perturbation near the boundary in the original domain. The effective range (for one $e$-folding) of the Green's function's response for mode $k$ is roughly $1/k$, and for the data used in this paper this value is roughly 10 cm for the first mode. Thus for a given point in the domain, density perturbation sources up to 10 cm away contribute significantly to the solution for the pressure perturbation at that point. This is a big reason why cropping the domain produces errors near the edges but not in the middle; the points near the edges are missing density perturbation sources from the cropping, while the points in the middle are mostly unaffected because they ``see'' all of their appropriate sources within the effective range. The extension that mimics the beam in the diffused buffer region provides approximate density perturbation sources for the points near the boundaries to reduce the error. Buffering the top and bottom of the domain does not have the same limitation as buffering the sides, and can be taken as large as one wants. However, for the data set presented here, not much was gained beyond $15\%$ buffering and the results do not seem to converge to the real answer near the edges for larger buffering, since the beam extension in the buffered region never quite looks like the original beam that has been cropped away.

\begin{figure*}
\includegraphics[width=\textwidth]{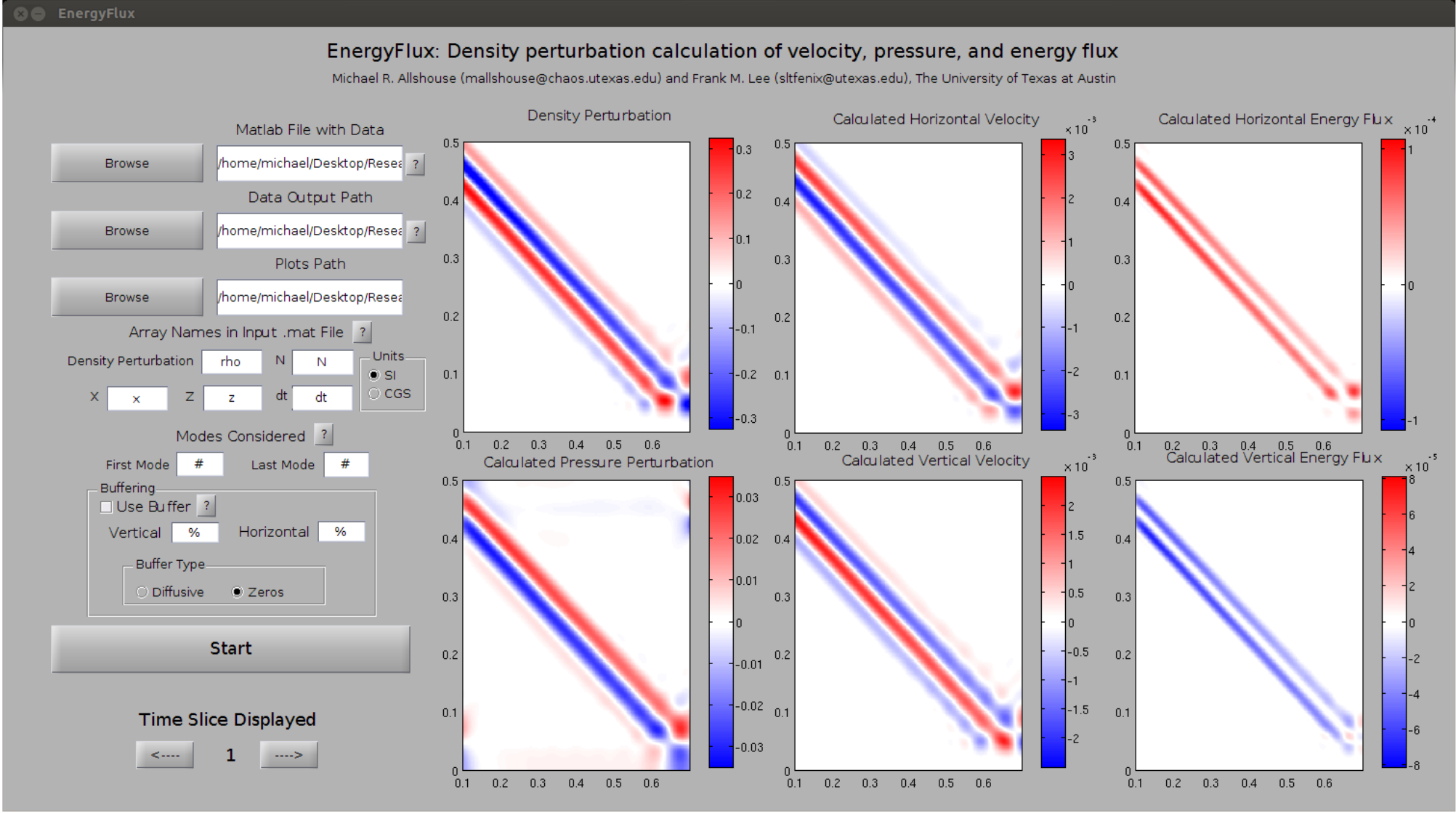}
\caption{Demonstration of the GUI interface {\it EnergyFlux} featuring the settings used for the results in section~\ref{subsec:Labdata}.}
\label{figA3}
\end{figure*}

\section{Implementation of Matlab GUI {\it EnergyFlux}}
\label{sec:gui}

To aid in the implementation of this method, a graphical user interface {\it EnergyFlux} was developed for Matlab.  This software is available in the supplemental materials along with a tutorial for use.  The GUI requires only the density perturbation field over a number of time steps, the corresponding coordinates, buoyancy frequency, and time step size.  The GUI allows for the implementation of the data buffering procedure presented in appendix~\ref{sec:cropped} and the selection of what range of horizontal modes to consider in the calculation.

\bibliographystyle{natbib}

\end{document}